\documentclass[pra,twocolumn,showkeys]{revtex4-1}
\usepackage{amsmath,amssymb,mathrsfs}
\usepackage{graphicx,import}
\usepackage[usenames,dvipsnames]{color}
\usepackage[colorlinks=true,linkcolor=blue,citecolor=BrickRed,urlcolor=blue]{hyperref}

\def\be{\begin{equation}}      
\def\ee{\end{equation}}
\def\bea{\begin{eqnarray}}      
\def\eea{\end{eqnarray}}

\begin{document}

\title{Anatomy of a periodically driven $p$-wave superconductor}

\author{Erhai Zhao}
\affiliation{Department of Physics and Astronomy, George Mason University, Fairfax, VA 22030}

\begin{abstract}
The topological properties of periodically driven many-body systems often have no static analogs and defy a simple description based on the effective Hamiltonian. To explore the emergent edge modes in driven $p$-wave superconductors in two dimensions, we analyze a toy model of Kitaev chains (one-dimensional spinless $p$-wave superconductors with Majorana edge states) coupled by time-periodic hopping. We show that with proper driving, the coupled Kitaev chains can turn into a fully gapped superconductor which is analogous to the $p_x+ip_y$ state but has two, rather than one, chiral edge modes. A different driving protocol turns it into a gapless superconductor with isolated point nodes and completely flat edge states at quasienergy $\omega=0$ or $\pi/T$, with $T$ the driving period. The time evolution operator $U(k_x,k_y,t)$ of the toy model is computed exactly to yield the phase bands. And the ``topological singularities" of the phase bands are exhausted and compared to those of a periodically driven Hofstadter model which features counter-propagating chiral edge modes. These examples demonstrate the unique edge states in driven superconducting systems and suggest driving as a potentially fruitful route to engineer new topological superconductors.
\end{abstract}
\maketitle

Among the best known examples of topological superconductors and superfluids \cite{RevModPhys.82.3045,RevModPhys.83.1057}, quite a few happen to have $p$-wave orbital symmetry. These include the Kitaev chain \cite{kitaev2001unpaired}, a spinless $p$-wave superconductor in one dimension (1D) with Majorana edge states at zero energy, the spinless $p_x+ip_y$ superconductor in two dimensions (2D) with chiral edge modes \cite{PhysRevB.61.10267}, and the B phase of superfluid Helium three \cite{vollhardt1990superfluid} in three dimensions (3D) with Majorana fermion surface states \cite{volovik2009universe,PhysRevLett.103.235301,PhysRevB.88.184506}. According to the general classification scheme \cite{PhysRevB.78.195125,kitaev}, these three fully gapped topological phases 
inhabit a diagonal line in the periodic table of topological insulators and superconductors. In parallel, it is also well known that unconventional superconductors with gapless bulk spectra can have nontrivial topological properties too. For example, the A phase of superfluid Helium three has a pair of stable point nodes \cite{vollhardt1990superfluid,volovik2009universe}, and accordingly, Fermi arc surface states \cite{PhysRevB.83.094510}. The edge states of these static (time-independent) superconductors are related to the bulk topological invariants via the bulk-boundary correspondence and well understood.

In this paper, we examine the edge states and topological properties of a model two-dimensional $p$-wave superconductor under time-periodic driving of various forms. We show that new types of edge states arise in these so-called Floquet superconductors. They cannot fit exactly into any known classes of static topological superconductors. In fact it is not always feasible to describe their topological properties using a static effective Hamiltonian $\mathscr{H}$ alone: the singularities \cite{nathan} of the time evolution operator $U(t)$ obstruct the smooth deformation of $U(t)$ into the form $\exp [{-i\mathscr{H} t}]$. Under these circumstances, one has to analyze $U(t)$ for the whole period of driving and from it construct the topological invariants. The unique features of Floquet topological insulators  \cite{cayssol2013floquet} and superconductors have been recognized by many authors (for a brief review see section I). The time-periodic (Floquet) states of $p_x+ip_y$ superconductor following a quench have also been studied \cite{PhysRevA.92.053620,PhysRevLett.113.076403,PhysRevB.88.104511}. Yet a complete classification of topological phenomena in periodically driven systems, to our best knowledge, is still lacking.

The main motivation of our work is to find perhaps the simplest possible examples of periodically driven superconductors in two dimensions that can be related to the familiar $p$-wave superconductors mentioned in the opening paragraph. To this end, we base our model on coupled Kitaev chains \cite{kitaev2001unpaired} but introduce the new ingredient of time-dependent hopping. The model is simple enough to be analytically tractable. Meanwhile it allows rich possibilities including gapped or gapless topological phases, and accordingly, edge states such as Majorana zero modes, chiral edge modes, and flat edge states. We carry out detailed analysis of the phase bands to reveal the point, line and plane degeneracies in the space of $(k_x,k_y,t)$ which play a key role in understanding the topology of the unitary time evolution. These examples provide clear evidence that there is much more to periodically driven systems than the effective Hamiltonians. And the Kitaev chains in particular, once allowed to talk to each other dynamically, turn into a bustling place for Floquet topological superconductivity.

\section{Topological singularities in periodically driven systems}

Recent theoretical work has firmly established that periodic driving can make an ordinary insulator or superconductor topologically nontrivial \cite{kitagawa_topological_2010,lindner_floquet_2011,jiang_majorana_2011,PhysRevB.87.235131,gu_floquet_2011,kitagawa_transport_2011,dora_optically_2012,PhysRevLett.110.016802,PhysRevLett.111.047002,cayssol2013floquet,PhysRevLett.111.136402,PhysRevB.88.155133,PhysRevB.87.201109,rudner_anomalous_2012,PhysRevA.89.063628,PhysRevLett.112.026805,PhysRevB.90.205108,PhysRevB.90.195419,nathan,PhysRevA.91.063626,PhysRevB.91.214518,epl15,PhysRevLett.113.266801,PhysRevB.89.121401}. Roughly speaking, a proper driving field mixes the bands to fundamentally change their topological characteristics such as the Berry curvature and Chern number \cite{lindner_floquet_2011}. For fast driving, the driven system stroboscopically mimics a static system described by the effective Hamiltonian which can be computed in a controlled manner by successive approximations.  Experimentally, Floquet edge states have been demonstrated in photonic crystals \cite{rechtsman_photonic_2013} and photonic quantum walks \cite{kitagawa_observation_2012}. In particular, time periodically modulated, or shaken, optical lattices \cite{eckardt_superfluid-insulator_2005,martin} have been implemented experimentally to engineer the band structure and generate artificially gauge fields for ultracold atoms \cite{PhysRevLett.95.170404,PhysRevLett.99.220403,PhysRevLett.102.100403,struck2011quantum,struck_tunable_2012,hauke_non-abelian_2012,struck_engineering_2013,parker2013direct,PhysRevLett.114.055301,jotzu2014experimental}. Given the scarcity of available topological superconductors with desirable properties, e.g., Majorana zero modes useful for topological quantum computing, it is desirable to explore to what extend periodic driving can be used to synthesize new topological superconductors using existing materials. This approach may be referred to as ``Floquet engineering."

The topological properties of periodically driven systems are complex and some of their features are rather unique. Due to the discrete time translational invariance of the Hamiltonian, the quasienergy spectrum of the driven system lives in the Quasienergy Brillouin Zone (QBZ) which is topologically equivalent to a closed circle. For a typical spectrum with $q$ bands, there are $q$ gaps, one more than the analogous static system. It can therefore support more edge modes, e.g. the so-called $\pi$-modes inside the gap around  quasienergy $\pm \pi/T$. Floquet Majorana modes at $\pi/T$ in one-dimensional systems were noted by many groups, for example in Ref. \cite{jiang_majorana_2011}. For 2D lattice systems, the $\pi$-modes have been studied by Rudner et al. \cite{rudner_anomalous_2012}, and in the context of periodically driven Hofstadter model by us \cite{PhysRevLett.112.026805,PhysRevB.90.205108} and also in Ref. \cite{PhysRevB.90.195419,PhysRevE.93.022209,PhysRevB.93.075405}. Kitagawa et al. have shown that the Floquet operator $U(T)$ can be used to construct the topological invariants for driven lattice systems in one and three dimensions \cite{kitagawa_topological_2010}. The topological invariants for 2D driven lattice systems have been constructed by Rudner et al. \cite{rudner_anomalous_2012} and Carpentier et al. \cite{PhysRevLett.114.106806}.

In the rest of the section, we review some of the concepts and definitions relevant to our subsequent discussion on driven superconductors. Consider a many-body system described by time-periodic Hamiltonian $H(t)=H(t+T)$, with $T$ the period. The time evolution operator 
\be
U(t)=\mathcal{T}e^{-i\int_0^t {H}(t') dt'},
\ee 
where $\mathcal{T}$ denotes the time ordering, $\hbar=1$, and we choose $U(t=0)=1$. Following the convention in the literature, we will call $U(t=T)$ the Floquet operator. The eigenvalue problem of $U(T)$ defines quasienergy $\omega_\ell$,
\be
U(T)|\psi_\ell\rangle=e^{-i\omega_\ell T}|\psi_\ell\rangle,
\ee
where $\ell$ is the band index, $\omega_\ell$ is equivalent to  $\omega_\ell+m\Omega$, with $m$ an integer and the fundamental frequency $\Omega=2\pi/T$. The first Quasienergy Brillouin Zone (QBZ) is usually defined as $\omega\in [-\Omega/2,\Omega/2]$. 
The effective Hamiltonian $\mathscr{H}$, which is time-independent, is defined from $U(T)$ through the relation 
\be
U(T)=e^{-i\mathscr{H} T}.
\ee

In order to understand the topological properties of the driven system, $\mathscr{H}$ and $U(T)$ are, generally speaking, insufficient. One often needs the entire function of $U(t)$ within a driving cycle, e.g. for $t\in [0, T]$. Following Nathan and Rudner \cite{nathan}, we introduce the notion of phase bands. Let $\mathbf{k}=(k_x,k_y)$ be the crystal momentum of a 2D lattice system. For any given time $t$, the eigenvalue problem of $U(\mathbf{k},t)$ yields an instantaneous band structure $\phi_\ell(\mathbf{k},t)$,
\be
U(\mathbf{k},t)|\phi_\ell(\mathbf{k},t)\rangle=e^{-i\phi_\ell(\mathbf{k},t)}|\phi_\ell(\mathbf{k},t)\rangle,
\ee
where the phase $\phi_\ell\in[-\pi,\pi]$. In particular, the phase bands at $t=T$ are nothing but the quasienergy bands, $\phi_\ell(\mathbf{k},t=T)=\omega_\ell(\mathbf{k}) T$. One can visualize the phase bands as a set of ``membranes" hovering above the 2D Brillouin Zone (BZ). As time goes on, these membranes change their shapes smoothly, and quite often during the process they intersect with or touch each other. These locations are where, to use a loose analogy, the ``knots are tied." It is illuminating to compare the actual evolution of the phase bands to that of a hypothetical static system described by the effective Hamiltonian $\mathscr{H}$ above,
\be
\mathscr{U}(\mathbf{k},t)=e^{-i\mathscr{H}(\mathbf{k}) t},
\ee
whose phase bands are simply given by $\omega_\ell(\mathbf{k}) t$, i.e., linearly dispersing with time. As shown by Nathan and Rudner \cite{nathan}, the scrambling of the phase bands of $U$ during the time evolution may render it topologically distinct from the linear evolution according to $\mathscr{U}$. Namely, it may be impossible to smoothly deform one to the other,
\be
U(\mathbf{k},t) \not\Leftrightarrow \mathscr{U}(\mathbf{k},t).
\ee
It then follows that the degeneracies in the phase bands, i.e. when and where the phase bands touch each other, hold the key to understand the topology of $U(\mathbf{k},t)$ and the corresponding edge states. From this argument, it is also clear that the effective Hamiltonian $\mathscr{H}$ by itself {\it cannot} provide a complete description of the topological properties of the driven system in all cases \cite{nathan}. Otherwise, it would have implied that the time evolution is always topologically equivalent to $\mathscr{U}$ and in turn equivalent to a static system. The presence of degeneracies in the phase band obstructs the deformation from $U$ to $\mathscr{U}$ and gives rise to edge modes unique to periodically driven (Floquet) systems. 

The knottiness of $U(\mathbf{k},t)$ can be captured by constructing its topological invariants. One example is the winding number introduced by Rudner et al. for 2D lattice systems \cite{rudner_anomalous_2012}. For the $\ell$-th gap of the quasienergy spectrum,
\begin{align}
w_\ell=\int \frac{dk_xdk_ydt}{24\pi^2} \epsilon^{\mu\nu\rho}\mathrm{Tr} \left[(u^{-1}\partial_\mu u) 
(u^{-1}\partial_\nu u)
 (u^{-1}\partial_\rho u)\right].
\end{align}
Here the greek indices loop through $k_x,k_y,t$ and are summed over. Note that $u(\mathbf{k},t)$
is an extrapolation of $U(\mathbf{k},t)$, for example \cite{rudner_anomalous_2012}, 
\be
u(\mathbf{k},t)=U(\mathbf{k},2t)\Theta(\frac{T}{2}-t)+e^{-i2\mathscr{H}_\ell(\mathbf{k})(T-t)}\Theta(t-\frac{T}{2}),
\ee
where $\Theta$ is the step function. The dependence of $u(\mathbf{k},t$ on $\ell$ is through the definition of $\mathscr{H}_\ell=-(i/T)\log U(T)$ where the branch cut of the logarithm is chosen to lie within the $\ell$-th gap  \cite{rudner_anomalous_2012}. The second term was added to unwind the evolution due to $\mathscr{H}$ and ensure $u(T)=1$. Via the bulk-boundary correspondence, the authors of Ref. \onlinecite{rudner_anomalous_2012} showed that $w_\ell$ is equal to the net chirality $\nu$ of the edge modes within the $\ell$-th gap, i.e. the number of chiral edge modes with positive group velocity minus the number of chiral edge states with negative group velocity.

To identify the features of the phase bands that give rise to a nonzero winding number, we review the argument given by Nathan and Rudner \cite{nathan}. Higher dimensional degeneracy manifolds of the phase band (such as surfaces and lines) may be shrunk to isolated points or entirely eliminated by introducing additional perturbations to lift the degeneracy. For 2D lattice systems, the only topologically stable degeneracies appear to be isolated points in the space of $(k_x,k_y,t)$. These band touching points are known as Weyl points in the study of semimetals \cite{wan2011topological} and nodal superconductors \cite{PhysRevB.86.054504,weyl-preprint}, or diabolical points in a more general context. As well known, a Weyl point can be viewed as a magnetic monopole \cite{volovik2009universe} with topological charge $q=1$ or $-1$.  Now imagine the phase bands are smoothly deformed to become flat at $\phi=0$ except for the neighborhoods of these point degeneracies and a small time interval of linear evolution which does not contribute to $w_\ell$ \cite{nathan}. 
Consider the quasienergy gap at the boundary of QBZ, and suppose the phase bands above and below (noting that the QBZ is periodic) touch each other during the time evolution at a few isolated degeneracy (Weyl) points. Let $q_i$ be the topological charge of the $i$-th degeneracy point. Their total charge $Q=\sum_i q_i$ is a topological invariant. Hereafter we will follow Ref. \cite{nathan} and refer to these degeneracy points as ``zone edge singularities." The winding number of the $\ell$-th quasienergy gap is then given by ~\cite{nathan}
\be
w_\ell=\sum_{n\leq \ell}C_n-Q,
\label{qq}
\ee
where $C_n$ is the Chern number of the $n$-th quasienergy band. In particular, for the gap at the zone edge, the Chern number sum over all the bands will vanish, and $w_\ell$ is nothing but $-Q$. Thus, finding the winding number or the net chirality for the zone-edge gap is reduced to counting the total charge of the corresponding point (Weyl) singularities. This is a very simple but useful result. One of the goals of our paper is to provide  concrete examples to illustrate the topological singularities in the phase bands, which make the driven system interesting and distinct from static systems.

Why do we care about the topological singularities in the phase band if we already know how to evaluate the topological invariant $w_\ell$ from $U(t)$? The first reason is that $w_\ell$ contains only the total charge $Q$. In contrast, the complete list of $\{q_i\}$ and their locations, the ``charge distribution map" (see Fig. 1 below for example), obtained from the phase band analysis contain much more information. Imagine a scenario that, due to additional symmetries, the degeneracy points always come in {pairs of opposite charge}. Then $Q$ will be identically zero. Vanishing $Q$ or $w_\ell$, however, does not mean there are no robust edge states. We have previously shown that this occurs in the periodically driven Hofstadter model  \cite{PhysRevLett.112.026805,PhysRevB.90.205108}. For the gap at quasienergy $\pi/T$, the winding number is zero but there can be pairs of counter-propagating edge modes. This model will be analyzed briefly in section II.

The second reason why topological singularity is such a useful concept has to do with Floquet systems with gapless quasienergy spectra. Previous theoretical works have largely focused on fully gapped Floquet topological insulators/superconductors. But gapless Floquet systems can also be topologically nontrivial with interesting edge states. The band flattening procedure and the winding number mentioned above become ill defined when the quasienergy gap closes, say, at isolated points in $\mathbf{k}$ space. We will provide an example in section V to show that even in these cases, understanding the degeneracies in the phase band helps identify the topological invariants and the corresponding edge states. 

Despite being very useful, the phase band analysis may be a messy business. First of all, degeneracies are ubiquitous in the phase bands of periodically driven systems even for topological trivial cases. Secondly, as mentioned above, the manifolds of degeneracy may be of finite dimensions in the form of lines or surfaces. It is theoretically plausible that the continuous degeneracy can be either lifted or reduced to isolated points by perturbations. In practice, however, it remains a nontrivial task to perform such topological surgeries numerically for arbitrary phase bands. Thirdly, the degeneracies of all phase bands for the entire driving cycle $t\in[0,T]$ have to be exhausted and classified. And finally, in some cases (see example in Section IV), direct application of Eq. \eqref{qq} is impossible, either because the quasienergy bands are overlapping (so that Chern numbers are ill-defined unless additional perturbations are introduced) or $Q$ itself is ill-defined, e.g., when the spectrum is gapless at the zone edge. In Section IV, we will show an example how such difficulty can be overcome by generalizing Eq. \eqref{qq}. We hope that our case studies presented here can stimulate further application of the phase band analysis to other Floquet systems. 

\section{Example: monopoles in the phase band}

To illustrate the topological singularities in the phase band, we first consider the example of periodically driven Hofstadter model \cite{hofstadter_energy_1976,PhysRevLett.111.185302,PhysRevLett.111.185301} at flux 1/3. We have previously shown that driving gives rise to a series of phases with robust counter-propagating edge modes at quasienergy $\pi/T$ \cite{PhysRevLett.112.026805,PhysRevB.90.205108}. Here we further show that the degeneracy points in the phase bands of this model can be obtained analytically, and they take the form of Weyl points, or magnetic monopoles, in the space of $(k_x,k_y,t)$. Thus this model can serve as a clean-cut example of the topological singularity in periodically driven systems.

Consider spinless fermions hopping on the square lattice in the presence of a uniform magnetic field of 1/3 flux quantum per plaquette. We work in the Landau gauge in which the magnetic BZ is given by $k_x\in[-\pi/3,\pi/3]$, $k_y\in[-\pi,\pi]$ (the lattice spacing is set to one). The hopping amplitudes are modulated periodically in time: for $0<t<T_1$, there is only $x$ hopping so the Hamiltonian in $\mathbf{k}$ space
\be
H_x(\mathbf{k})=-J_x\left[
\begin{array}{ccc}
  0&  e^{-ik_x} &   e^{ik_x} \\
  e^{ik_x}& 0  &  e^{-ik_x} \\
  e^{-ik_x}&  e^{ik_x} & 0  
\end{array}
\right],
\ee
while for $T_1<t<T$, only the $y$ hopping is turned on,
\be
H_y(\mathbf{k})=-2J_y\left[
\begin{array}{ccc}
\cos k_y&  0 &   0 \\
  0& \cos(k_y+\frac{2\pi}{3}) &  0 \\
  0&  0 & \cos(k_y+\frac{4\pi}{3})  
\end{array}
\right].
\ee
In other words, we have $H(\mathbf{k},t+T)=H(\mathbf{k},t)$ and for $0<t<T$,
\be
H(\mathbf{k},t)=H_x(\mathbf{k}) \Theta(T_1-t) + H_y(\mathbf{k})  \Theta(t-T_1),
\ee
where $\Theta$ is the step function. It follows that the phase bands for $t<T_1$ are just three $\cos k_x$ bands shifted away from each other by $2\pi/3$,
\be
\phi_\ell(\mathbf{k},t)=-2\cos(k_x+\delta_\ell) J_xt,\;\;\delta_\ell=(\ell-1)\frac{2\pi}{3},\ell =1,2,3.
\ee
The cosine bands cross for example at $k_x=0$ and $\pm \pi/3$. Such degeneracies occur regardless of $k_y$ and $t$, therefore they are actually degeneracy {\it planes}. As soon as $J_y$ is turned on, $t>T_1$, these degeneracies are lifted,
and the phase bands become fully gapped and well separated from each other (for parameters corresponding to the phase B of Ref. \onlinecite{PhysRevLett.112.026805}). Only at some special points, the bands touch each other. These degeneracy {\it points} can be obtained analytically by examining the time-evolution operator $U(t>T_1)$. We will illustrate this below using the example of $\theta_x\equiv J_xT_1=\pi/3$.

\begin{figure}
\includegraphics[width=0.34\textwidth]{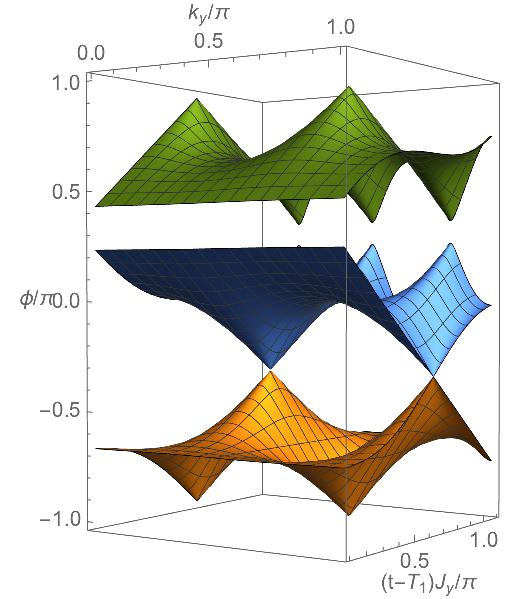}\includegraphics[width=0.16\textwidth]{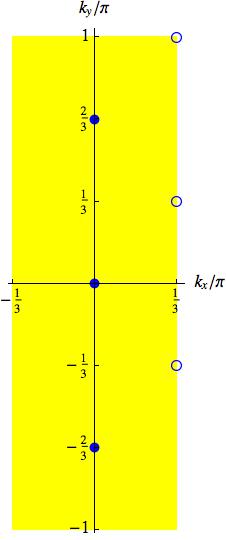}
\caption{Topological singularities in the phase bands of a periodically driven Hofstadter model at fixed flux 1/3. $k_x=0$, $\theta_x=\pi/3$.
The panel on the right shows the locations of the zone edge singularity with monopole charge $+1$ (empty circle) or $-1$ (solid dot) for
$t=T_1+2\pi/3J_y$. 
}
\label{qh}
\end{figure}

Let us first consider $k_y=0$, for which $H_y$ becomes a constant matrix. For the special case of $\theta_y\equiv J_y(t-T_1)=2\pi/3$, $\exp[-iH_y(t-T_1)]$ reduces to the identity matrix times $e^{i4\pi/3}$, and
\be
U(\mathbf{k},t>T_1)=e^{i4\pi/3} e^{-iH_x T_1}.
\ee
From this we can read out the phase bands,
\be
\phi_\ell =- [4\pi/3 + 2\theta_x \cos(k_x+\delta_\ell)].
\ee
Hereafter the values of $\phi$ are understood as modulo $2\pi$ and within $[-\pi, \pi]$. The three phase bands cross at $k_x=0, \pm \pi/3$. Similarly, for $k_y=\pi/3$ and $\theta_y=2\pi/3$, we have
\be
\phi_\ell =- [2\pi/3 + 2\theta_x \cos(k_x+\delta_\ell)],
\ee
which has the same crossing structure. Indeed all the degeneracy points of the phase band (relevant to phase B) are given by the following list,
\begin{align*}
&k_x=0,\pm \pi/3.\\ 
&k_y=0,\pm\pi/3,\pm 2\pi/3,\pm \pi. \\
&t=T_1+2\pi/3J_y.
\end{align*}
The left panel of Fig. \ref{qh} shows four isolated degeneracy (Weyl) points for fixed $k_x=0$. Two of them (at $k_y=\pi/3$ and $\pi$) are crossings of the bottom band $\ell=1$ and the middle band $\ell=2$. The remaining two (at $k_y=0$ and $2\pi/3$) are crossings at $\phi=\pi$ between the bottom band $\ell=1$ and the top band $\ell=3$. It is clear that a crossing of the latter kind is only possible in periodically driven systems: it occurs at the edge of QBZ, hence the name ``zone edge singularity" \cite{nathan}.  

We can exhaust the zone edge singularities for $t=T_1+2\pi/3J_y$ and the results are shown in the right panel of Fig. \ref{qh}. Within the magnetic BZ, there are three Weyl points with topological charge $+1$ (empty circles) and another three Weyl points with topological charge $-1$ (solid dots). The total topological charge $Q$ is zero, corresponding to a zero winding number and zero net chirality for the gap at the QBZ boundary, in agreement with the earlier analysis of Ref. \onlinecite{PhysRevLett.112.026805}. Vanishing $Q$ or winding number however does not mean that no edge states exist. On the contrary, we have shown in Ref.  \onlinecite{PhysRevLett.112.026805} that pairs of counter-propagating $\pi$-modes on the edge are robust. 

\section{Dynamically Coupled Kitaev Chains}

Our model of periodically driven $p$-wave superconductor consists of
spinless fermions hopping on the square lattice with pairing interaction between neighboring sites along the $x$ direction,
\be 
H=\sum_\mathbf{r}[-J_x c^\dagger_\mathbf{r} c_{\mathbf{r}+\hat{x}} - J_y c^\dagger_\mathbf{r} c_{\mathbf{r}+\hat{y}}+ \Delta c_\mathbf{r} c_{\mathbf{r}+\hat{x}} +h.c.]- \mu \sum_\mathbf{r} n_\mathbf{r}.
\label{dkc}
\ee
Here $\mathbf{r}$ labels the lattice sites and $n_\mathbf{r}=c^\dagger_\mathbf{r} c_{\mathbf{r}}$. We set the lattice spacing to be one and assume the hopping $J_x$ and $J_y$ can be independently controlled, for example, by varying the laser intensity of a square optical lattice.  We also assume that there is no pairing interaction between $\mathbf{r}$ and $\mathbf{r}+\hat{y}$. In the limit $J_y\rightarrow 0$, the system decouples into Kitaev chains, with each chain a 1D $p$-wave superconductor described by the Hamiltonian \cite{kitaev2001unpaired}
\be
H_x=\sum_i [-J_x c^\dagger_i c_{i+1} +\Delta c_i c_{i+1}+h.c.]-\mu\sum_i c^\dagger_i c_i,
\ee
where $i$ labels the lattice sites along the chain direction $\hat{x}$. For $|\mu|<2J$, the ground state is a fully gapped topological superconductor belonging to the BDI class \cite{PhysRevB.78.195125,kitaev}. Edge states form at the ends of an open chain exactly at zero energy. These fermionic quasiparticles are their own anti-particles, and therefore Majorana zero modes \cite{kitaev2001unpaired}. 
 
We focus on dynamical optical lattices where the hopping amplitudes in $H$ above are periodic functions of time, 
\be J_{x,y}(t)=J_{x,y}(t+T).
\ee
We shall call this model dynamically coupled Kitaev chains. Two specific examples of $J_{x,y}(t)$ will be analyzed below. While it is hard to control the hopping for solid state systems, dynamical optical lattices have been realized and studied by many groups \cite{PhysRevLett.95.170404,PhysRevLett.99.220403,PhysRevLett.102.100403,struck2011quantum,struck_tunable_2012,hauke_non-abelian_2012,struck_engineering_2013,parker2013direct,PhysRevLett.114.055301,jotzu2014experimental}. The central questions we try to address are as follows. (1) Is it possible to achieve a fully gapped Floquet superconductor by simply varying $J_{x,y}(t)$? (2) If so, what are the edge states and how do they differ from static topological superconductors? (3) Is it possible to achieve a gapless Floquet superconductor that is topologically nontrivial? (4) What will the time modulations do to the Majorana zero modes?

\section{Four-step driving}
\begin{figure}
\includegraphics[width=0.48\textwidth]{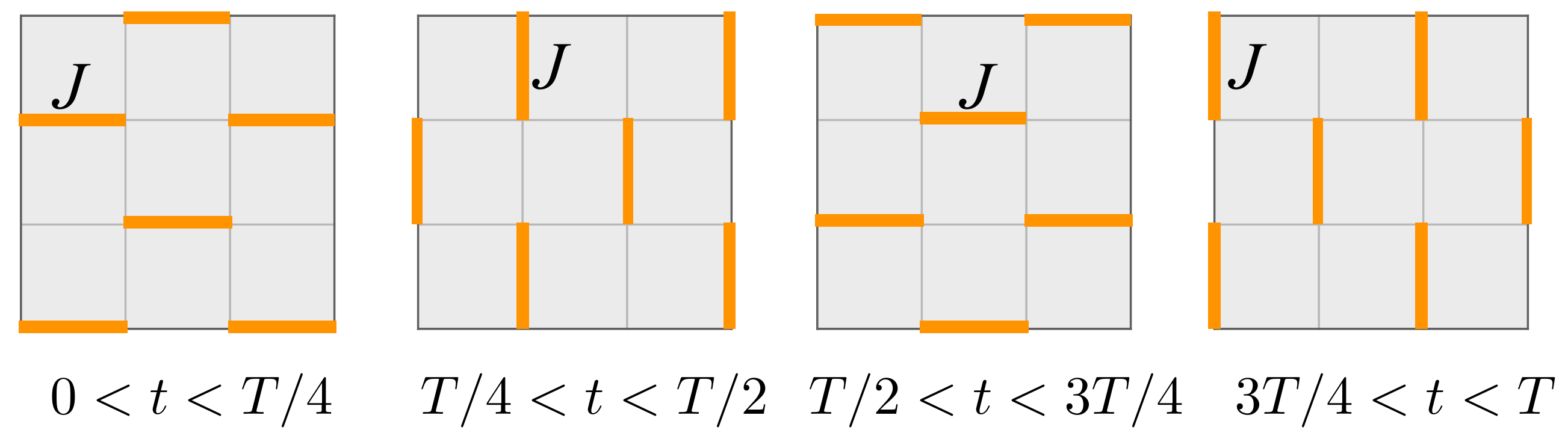}
\caption{The four-step driving protocol. The driving period is divided into four quarters. Within each quarter, only the hopping along the bonds labelled by orange thick lines are allowed. Other parameters such as $\mu$ and $\Delta$ are kept constant.}
\label{f4}
\end{figure}

In the first example, we employ a four-step driving protocol to prove that it is {\it possible} to turn the coupled Kitaev chains into a fully gapped topological superconductor with chiral edge modes similar to those of the $p_x+ip_y$ state. The four-step driving protocol is borrowed from the earlier work of Kitagawa et al. \cite{kitagawa_topological_2010} and Rudner et al. \cite{rudner_anomalous_2012} who studied 2D lattice models of Floquet topological insulators. As illustrated in Fig. \ref{f4}, during each quarter of the driving period $T$, e.g. $t\in [0,T/4]$, only hopping along $1/4$ of the bonds is turned on and set to $J$ (indicated by the orange thick lines) while the hopping along the rest of the bonds is identically zero. Intuitively, one anticipates that Majorana zero modes will cease to exist, because there is never a ``chain" at any given time. It is not immediately clear though whether there are any quasienergy gaps or edge states in the driven system.

\begin{figure}
\includegraphics[width=0.4\textwidth]{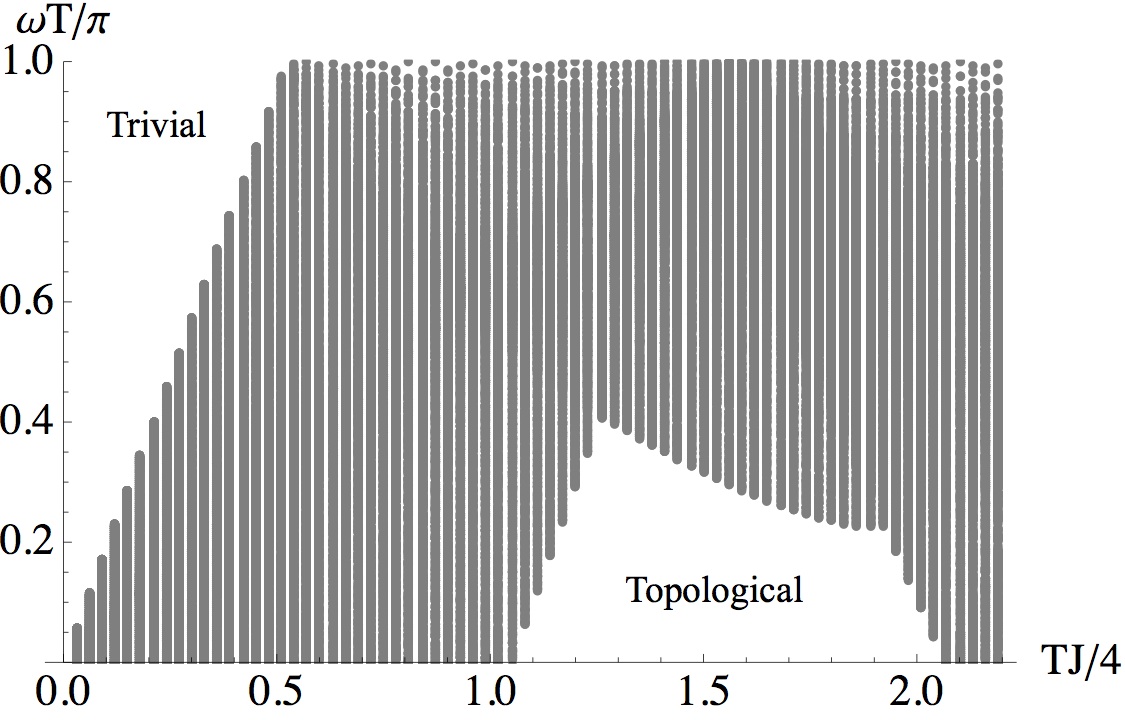}
\caption{Gaps (empty regions) in the quasienergy spectrum for the four-step driving protocol. Gray dots are $\omega(\mathbf{k})$ for $\mathbf{k}$ inside the reduced Brillouin zone. $\Delta=0.2J$, $\mu=-0.5J$. }
\label{f4gap}
\end{figure}

\begin{figure}
\includegraphics[width=0.24\textwidth]{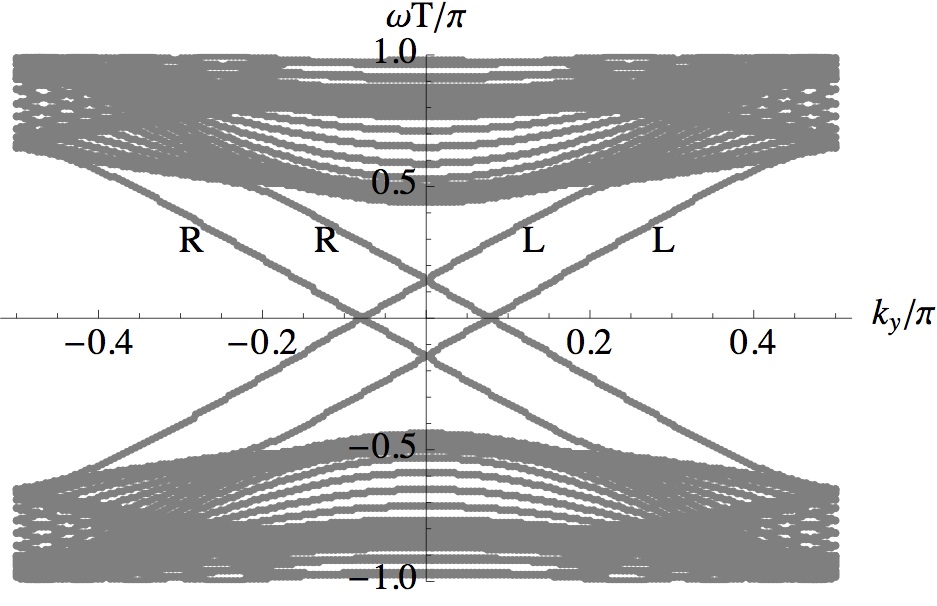}\includegraphics[width=0.24\textwidth]{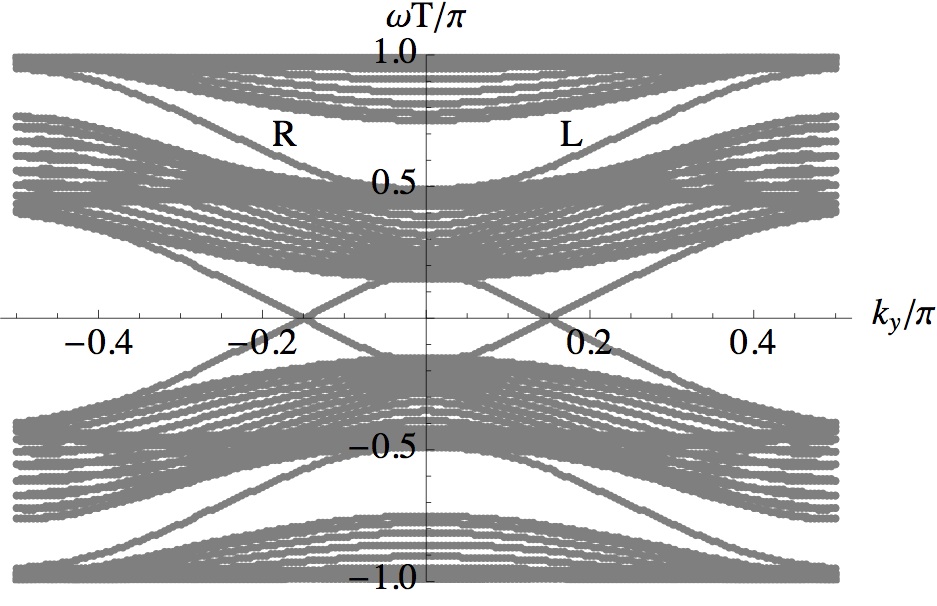}
\caption{The quasienergy spectrum of a slab of width $L_x=20$ with four-step driving. Left: $TJ/4=1.3$, $\mu=-0.5J$, and L (R) denotes states on the left (right) edge. Right: Introducing staggered potential $\mu_A=-0.25J$ and $\mu_B=-0.75J$ for the A and B sublattice opens additional gaps, 
$TJ/4=1.1$. $\Delta=0.2J$ in both cases.}
\label{f4edge}
\end{figure}

To find the bulk quasienergy spectrum, we compute the Floquet operator $U(t=T)=U_4U_3U_2U_1$, with $U_i=\exp [-i H_i T/4]$ and $H_i$ being the Hamiltonian within the $i$-th quarter of the driving period. We take $\Delta=0.2J$, $\mu=-0.5J$ as an example and plot the allowed quasienergies $\omega(\mathbf{k})$ for $\mathbf{k}$ inside the reduced Brillouin zone in Fig. \ref{f4gap} for varying driving period $T$. Because the spectrum has particle-hole symmetry, only $\omega\in [0,\pi/T]$ needs to be shown. Two gapped phases can be identified from Fig. \ref{f4gap}. The first phase occurs for $TJ/4<0.55$, i.e. for small $T$ and thus fast driving, and its gap centers at $\omega=\pm \pi/T$. This phase appears to be topologically trivial: there is no edge states associated with the gap. In contrast, the second gapped phase, with its gap centering at  $\omega=0$ for $TJ/4 \in [1.05, 2.05]$, seems topologically nontrivial: there are chiral edge modes inside the gap. To find the corresponding edge modes, we consider a slab of width $L_x=20$ in the $x$ direction but infinitely long along $y$. The quasienergy spectrum of the slab, shown in Fig. \ref{f4edge} for $TJ/4=1.3$, contains four edge states inside the bulk gap. By inspecting their wave functions directly, one finds that the two edge modes with the {\it same group velocity} are located at the {\it same edge}.

\begin{figure}
\includegraphics[width=0.4\textwidth]{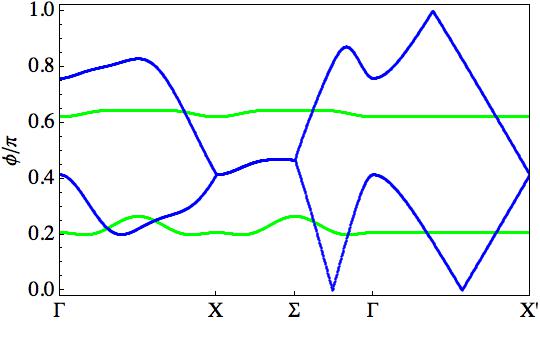}
\includegraphics[width=0.4\textwidth]{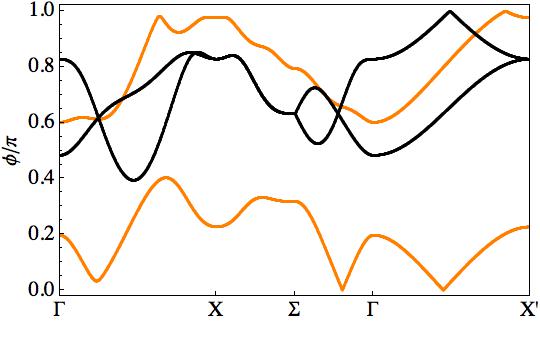}
\caption{The phase band for $t=T/4$ (green), $t=T/2$ (blue), $t=0.726T$ (orange), and $t=T$ (black, i.e. the quasienergy spectrum). $TJ/4=1.3$, $\Delta=0.2J$, $\mu=-0.5J$.}
\label{f4band}
\end{figure}

To summarize, time-modulation of the hopping amplitudes turns a lattice system with static $p_x$ pairing interaction into a Floquet superconductor whose edge states are analogous, but not identical, to those of the $p_x+ip_y$ state. The $p_x+ip_y$ superconductor only has one band (we only count the number of particle branches at positive energies). The driven system considered here has two overlapping quasienergy bands $\omega_{1,2}(\mathbf{k})>0$  and two chiral edge modes per edge. To shed more light on their topological properties, we generalize Eq. \eqref{dkc} by introducing different chemical potentials for the A and B sublattices. As shown in the right panel of Fig. \ref{f4edge}, $\omega_{1}$ and $\omega_{2}$ are now separated by a gap,
i.e., the bands are ordered as $0<\omega_1<\omega_2$. Inside the newly generated gap, there is also a chiral edge mode. The Chern number for the $\omega_{1}(\mathbf{k})$ band is $-1$, which is consistent with the fact that there is one edge mode stemming from it from above and two from below. Notice however that the spectrum still remains gapless at the zone edge and therefore $Q$ is ill-defined. Obviously Eq. \eqref{qq} cannot be directly applied here to predict the number of edge modes for the gap centered at $\phi=0$. A relation similar to Eq. \eqref{qq}  is needed.

\begin{figure}
\includegraphics[width=0.4\textwidth]{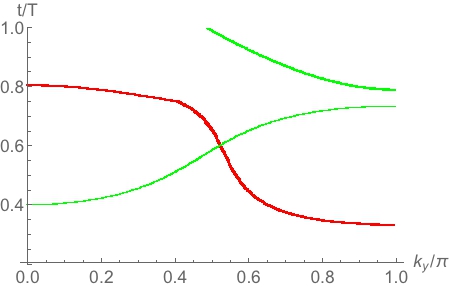}
\includegraphics[width=0.4\textwidth]{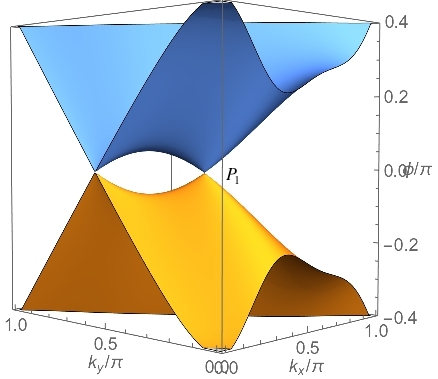}
\caption{Top: lines of degeneracies at $\phi=0$ (red) and $\phi=\pi$ (green) in the phase band for fixed $k_x=0$. Bottom: point degeneracy at $k_x$=$k_y$=$0.26\pi$, $t=0.5T$. $TJ/4=1.3$, $\Delta=0.2J$, $\mu=-0.5J$.
}
\label{f4line}
\end{figure}

To this end, it is instructive to inspect the phase bands, $\phi_\ell(\mathbf{k},t)\geq 0$ with $\ell=1,2$, as shown in Fig. \ref{f4band} for the topological phase at $TJ/4 =1.3$ and four representative time instants throughout the driving cycle $t=T/4$, $T/2$, $0.726T$ and $T$. Note that the BZ is a diamond because the unit cell consists of two lattice sites for the four-step driving. The spectrum does not obey $C_4$ rotational symmetry due to the $p_x$ pairing term. So we have included the cut from $\Gamma$ to $X'$ along the $k_y$ axis in the band structure. We observe that two phase bands are degenerate along the BZ boundary $X\Sigma$ for $t=T/2$ (blue curves) and $t=T$ (black curves). More importantly, as illustrated by the blue and orange curves in Fig. \ref{f4band}, the bands also touch at $\phi=0$ or $\pi$. Due to the particle-hole symmetry, this is where they touch their mirror reflected particle-hole conjugate bands (not shown). Such degeneracies at $\phi=0$ or $\pi$ appear in two ways, either as a continuous line or an isolated point in the space of $(k_x,k_y,t)$. The top panel of Fig. \ref{f4line} depicts the lines of degeneracies at $\phi=0$ (in red) and $\phi= \pi$ (in green) within the plane of $k_x=0$. There are also two Weyl-type point degeneracies at $\phi=0$ occurring at 
\[ 
P_1:\;\; (k_x=k_y=0.26\pi,\;t=0.5T),
\]
shown in the bottom panel of Fig. \ref{f4line}, and 
\[
P_2:\;\; (k_x=k_y=0.2\pi,\;t=0.726T).
\]
These two Weyl points are also visible in Fig. \ref{f4band} along the $\Sigma \Gamma$ cut in the blue and orange curves respectively. They are very similar to the Weyl points in the example of periodically driven Hofstadter model in section II.

The degeneracy line for $k_x=0$ can be traced back to the $p_x$ pairing term $\Delta\sin k_x$ which vanishes for $k_x=0$ and renders the phase band gapless at $\phi=0$.  We may argue that its stability is not topologically protected as follows. One can imagine a smooth deformation of $H_i$ that shrinks the degeneracy line to the $\Gamma$ point (or $X'$ point) where it annihilates with its conjugate partner with opposite momentum. Therefore, the line degeneracy can in principle be entirely lifted, and we only need to consider the two stable Weyl points $P_1$ and $P_2$. 
The topological charge of each Weyl point can be evaluated following the standard procedure. At the Weyl point, the spectrum is doubly degenerate. The two eigenstates $|\pm\rangle$ span a pseudospin space. The phase band spectrum deviating from the Weyl point in the $k_x$, $k_y$, $t$ direction can be projected onto the pseudospin space to obtain its three pseudospin components. The result is a $3\times 3$ matrix $S$ \cite{nathan,weyl-preprint}. The topological charge is simply 
\be
q=\mathrm{sign} [\det S]. 
\ee
In this way, we find both $P_1$ and $P_2$ carry unit topological charge. So their total charge $Q_0=q_1+q_2=2$.

Now we are ready to relate these topological charges to the number of edge modes at the $\phi=0$ gap. Instead of working with $U(t)$ defined above, it is more convenient to analyze the topological properties of 
\be
U_s(t)=U(t+t_0),
\ee
with $t\in [0,T]$ and $t_0$ is some small but finite value, e.g. $t_0=0.05T$, chosen so that the phase bands of $U_s(t=0)=U(t_0)$ are gapped at $\phi=0$ and topological trivial with vanishing Chern number [similar to the green curves in Fig. \ref{f4band}]. Note that the phase bands of $U_s(t=T)$ [similar to the black curves in Fig. \ref{f4band}] are also gapped at $\phi=0$. During the time evolution between $t=0$ and $t=T$, the gap at  $\phi=0$ collapses and reopens twice at $P_1$ and $P_2$ respectively. By similar arguments given in Ref. \onlinecite{nathan} for the zone edge singularities, we are led to the conclusion that the total number of chiral edge modes, $\nu_0$, is given by the total charge of $P_1$ and $P_2$,
\be
\nu_0=-Q_0=-2,
\label{nu0}
\ee
for the gap  centered at $\phi=0$. This agrees with the two chiral edge states obtained directly from finite slab calculation. The construction in Ref. \onlinecite{nathan} focuses on the zone-edge gap, which is open at $t=0$ and $t=T$, and exploits the fact that the sum of the Chern numbers of all the bands is zero. By focusing on $U_s(t)$, we fulfill the same requirements and ensure that the Chern number for all the bands at $t=0$ is zero. This enables us to draw the conclusion Eq. \eqref{nu0} without separating the quasienergy bands or computing the Chern numbers. Note that $U_s(t)$ amounts to choosing a starting time of the driving cycle, or a gauge choice of $U(t)$. 

Our toy model has overlapping phase bands and several kinds of degeneracies, both of which are common features of driven systems. Despite these complications, we are able to locate the topological singularities at the quasienergy gap, and relate them to the number of Floquet edge modes. The example also illustrates that driving can serve as a potentially fruitful way to achieve topological superconductivity: one does not need exotic pairing mechanisms provided that there is  sufficient control over the hopping patterns of fermions.

\section{Two-step driving}
In the second example, the driving only consists of two steps: first only the hopping along the $x$ direction is turned on, then it is switched off and the $y$ hopping is on, just like the driven Hofstadter model discussed in section II above. It is simpler than the four-step driving, and its time-evolution operator can be obtained analytically. The Hamiltonian for the dynamically coupled Kitaev chain, $H(\mathbf{k},t+T)=H(\mathbf{k},t)$, takes the following form for $t\in[0,T]$,
\be
H(\mathbf{k},t)=H_x(\mathbf{k}) \Theta(T_1-t) + H_y(\mathbf{k})  \Theta(t-T_1),\\
\ee
where $\Theta$ is the step function and
\begin{align}
H_x(\mathbf{k}) &=(-2J_x\cos k_x-\mu)\tau_z -2\Delta \sin k_x \tau_y,\\
H_y(\mathbf{k}) &=(-2J_y\cos k_y-\mu)\tau_z -2\Delta \sin k_x \tau_y.
\end{align}
We have introduced $\tau_i$ as the Pauli matrices in the Nambu (particle-hole) space. As a $2\times 2$ unitary matrix, $U(\mathbf{k},t)$ is a rotation in the Nambu space around some axis $\hat{d}(\mathbf{k},t)$ by an angle $\phi(\mathbf{k},t)$,
\be
U(\mathbf{k},t)=e^{i\phi(\mathbf{k},t) (\boldsymbol{\tau}\cdot \hat{d})}.
\ee
The two phase bands are simply $\pm\phi(\mathbf{k},t)$ and they obey the particle-hole symmetry. For  $t<T_1$, $U(t)=\exp[-iH_x t]$ to yield
\be
\phi(\mathbf{k},t)=\omega_1 t,\;\;\omega_1=\sqrt{(2J\cos k_x+\mu)^2+(2\Delta\sin k_x)^2}.
\ee
For $t>T_1$, $U(t)=\exp[-iH_y (t-T_1)]\exp[-iH_x T_1]$. These two successive rotations in the Nambu space can be combined using the standard formula of Pauli matrices to find $\phi(\mathbf{k},t>T_1)$. The expression is lengthy and will not be explicitly listed here.

\begin{figure}
\includegraphics[width=0.4\textwidth]{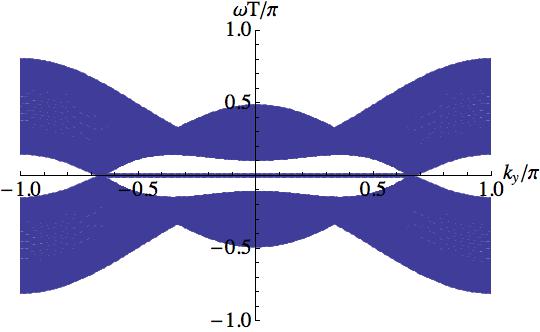}
\includegraphics[width=0.4\textwidth]{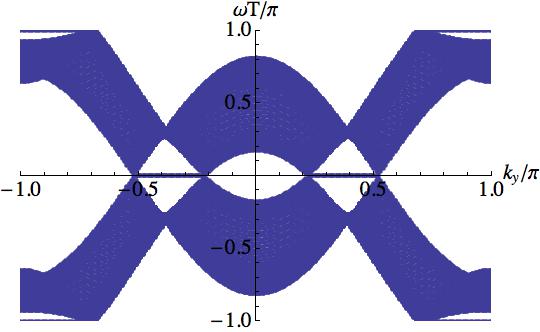}
\caption{The quasienergy spectrum of a slab under two-step driving for $(T-T_1) J=0.5$ (upper panel) and 1.2 (lower panel). $\Delta=-\mu/2=0.25J$, $T_1 J=0.5$.}
\label{fg2ed}
\end{figure}

To see how the quasienergy spectrum looks like under two-step driving, we fix the value of $\theta_x=J_xT_1$ to be $0.5$, and gradually increase $\theta_y=J_y(T-T_1)$, i.e. increasing the driving period and thus decreasing the driving frequency. For simplicity, we set $J_x=J_y=J$. For sufficiently large $\theta_y$, the dynamically coupled Kitaev chains become a gapless Floquet superconductor with one pair of point nodes. The upper panel of Fig. \ref{fg2ed} shows the quasienergy spectrum of a slab with width $L_x=40$ for $\theta_y=0.5$. We find a flat edge mode at zero energy connecting the two nodal points located at $k_y=\pm \kappa$. With even larger $\theta_y$, the bulk spectrum develops two pairs of nodal points, as shown in the lower panel of  \ref{fg2ed} for $\theta_y=1.2$. In this case, a line of flat edge mode at zero energy connects the nodal point at $k_y=\kappa$ and $\kappa'$, while another line of zero energy mode connects $-\kappa$ and $-\kappa'$. This suggests that bulk-boundary correspondence is at play here. We will prove it below that these edge states are indeed the manifestations of the topologically nontrivial bulk quasienergy band.

The spectrum in Fig. \ref{fg2ed} is reminiscent of a Weyl superconductor/superfluid in 3D such as the A phase of superfluid $^3$He. The two nodal points of $^3$He A are topologically stable and give rise to a continuous line of gapless surface states at zero energy known as Fermi arcs. The analogy however is not precise because Weyl nodes only occurs in 3D and our system is two-dimensional (albeit dynamical). It is also tempting to view the zero energy edge states here as the remnant of the Majorana zero modes of the uncoupled Kitaev chains. Once $J_y$ is turned on, generally speaking, the original Majorana zero modes tend to disperse along $k_y$ to acquire a nonzero energy. The fact that they ``survive" for a range of $k_y$ values, such as $k_y\in [-\kappa, \kappa]$, suggests that they are protected by a topological invariant. More importantly, the driven system also acquires a unique feature that has no static analog. In the lower panel of Fig. \ref{fg2ed}, we also observe flat edge modes at quasienergy $\pi/T$, the ``flat $\pi$-modes."

To understand all these features, we consider a particular limit of the two-step driving: $(T-T_1)\rightarrow 0$ while keeping $\theta_y=J_y(T-T_1)$ constant. In other words, $J_y$ is only turned on for a very short while, $J_y(t)=\theta_y \sum_n\delta (t-nT)$. The driving protocol in this limit is known as periodic kicking \cite{PhysRevLett.112.026805}. Due to the vanishing time interval $T-T_1$, the $\mu$ and $\Delta$ terms in $H_y$ do not contribute to $U$ and $\exp[-iH_y (T-T_1)]$ reduces to $\exp [i2\theta_y \cos k_y \tau_y]$, i.e. a pure $\tau_y$ rotation in the Nambu space. Despite these idealizations to simplify the algebra, all the essential physics shown in Fig. \ref{fg2ed} is retained in the periodic kicking limit. The phase band at the end of the two-stage time evolution is given by  
\be
\cos\phi(\mathbf{k},T)=\cos \rho_2 \cos\rho_1
-\frac{2J\cos k_x+\mu}{\omega_1}\sin \rho_2\sin \rho_1,
\ee
where  
\be
\rho_1=\omega_1T, \;\; \rho_2=2\theta_y\cos k_y.
\ee
The components of the $\mathbf{d}$ vector are found to be 
\begin{align}
d_x=& \frac{2\Delta\sin k_x}{\omega_1}\sin \rho_2 \sin \rho_1, \\
d_y=& \frac{2\Delta\sin k_x}{\omega_1} \cos \rho_2 \sin \rho_1, \\
d_z=& \sin \rho_2 \cos \rho_1 + \frac{2J\cos k_x+\mu}{\omega_1} \cos \rho_2 \sin \rho_1.
\end{align}
The quasienergy $\omega(\mathbf{k})=-\phi(\mathbf{k},T)/T$ and 
the effective Hamiltonian  $\mathscr{H}=\omega(\mathbf{k}) (\boldsymbol{\tau}\cdot \hat{d})$. 

\begin{figure}
\includegraphics[width=0.35\textwidth]{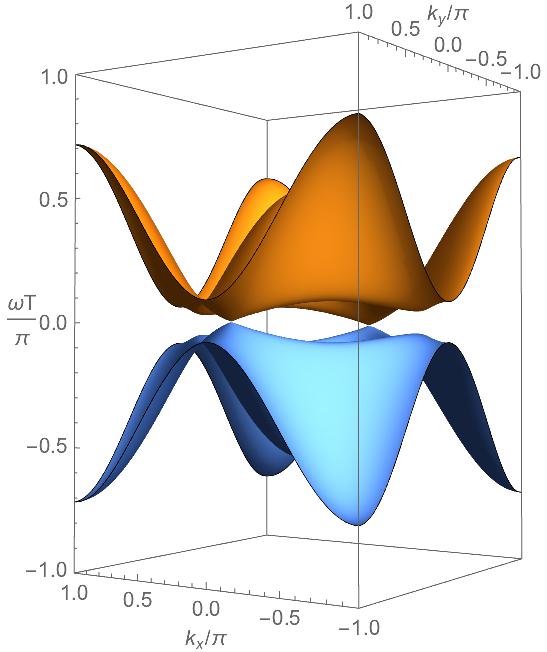}
\includegraphics[width=0.35\textwidth]{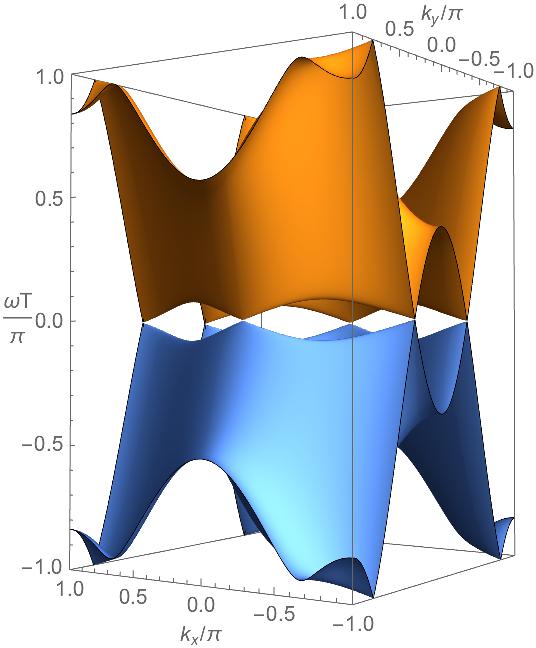}
\caption{The bulk quasienergy band $\omega(\mathbf{k})T$ for the two-step driving in the periodic kicking limit, $\theta_y=0.5$ (upper panel) and 1.2 (lower panel). $\Delta=-\mu/2=0.25J$, $T J=0.5$.
}
\label{fg2bulk}
\end{figure}

As before, we fix the value $\theta_x=JT=0.5$ and gradually increase $\theta_y$ for the periodically kicked system. The quasienergy bands $\omega (\mathbf{k})$ undergo a series of phase transitions at critical points $\theta_{c1}< \theta_{c2}<\theta_{c3}<...$ (here we use the term ``phase" very loosely to refer to a region of the parameter space where the bulk and edge spectra of the driven system develop distinctive characteristics). For small $\theta_y$, the bands are fully gapped and do not exceed the QBZ. The system remains topologically identical to the decoupled Kitaev chains. But as soon as $\theta_y$ exceeds $\theta_{c1}$, the gap at $\omega=0$ closes. This occurs at {a pair of nodal points} in $\mathbf{k}$ space with opposite momenta, $(k_x,k_y)=(0,\kappa)$ and $(0,-\kappa)$. The value of $\theta_{c1}$ can be obtained  by setting $k_x=0$, $k_y=\pi$ and requiring $\omega$ to be zero,
\be
\theta_{c1} = (2J+\mu)T.
\ee
As $\theta_y$ is increased beyond $\theta_{c1}$, $\kappa$ decreases continuously from $\pi$ along the $k_y$ axis. The pair of point nodes in the phase band are illustrated in the upper panel of Fig. \ref{fg2bulk} for the example of $\theta_y=0.5$.  Beyond a second critical point, $\theta_y\geq \theta_{c2}$, {another pair of point nodes} appear at momenta $(k_x,k_y)=(\pi,\kappa')$ and $(\pi,-\kappa')$, as shown in the lower panel of Fig. \ref{fg2bulk} for $\theta_y=1.2$. The value of $\theta_{c2}$ can be found by setting $k_x=\pi$, $k_y=0$ and requiring $\omega$ to vanish,
\be
\theta_{c2} = (2J-\mu)T.
\ee
As $\theta_y$ increases beyond $t_{c2}$, $\kappa'$  grows in magnitude. For even larger $\theta_y>\theta_{c3}$, 
the quasienergy bands also touch each other at the QBZ boundary $\omega=\pi/T$ as illustrated in Fig. \ref{fg2bulk} (lower panel). Again, this occurs at a pair of degeneracy points at $(k_x,k_y)=(\pi,\kappa'')$. By setting $k_x=\pi$, $k_y=\pi$ and requiring $\omega=\pi/T$, we find
\be
\theta_{c3}=\pi/2 - \theta_{c2}.
\ee
To summarize, from the bulk spectrum alone we can identify a series phases for fixed $\theta_x=0.5$: 
(1) Phase A, $\theta_y\in [0,\theta_{c1}]$, fully gapped;
(2) Phase B, $\theta_y\in [\theta_{c1},\theta_{c2}]$, with one pair of point nodes at $\omega=0$;
(3) Phase C, $\theta_y\in [\theta_{c2},\theta_{c3}]$, with two pairs of point nodes at $\omega=0$;
and (4) Phase D, $\theta_y\in [\theta_{c3},\theta_{c4}]$, with two pairs of point nodes at $\omega=0$, plus one pair of point degeneracy at $\omega=\pi$. In passing we mention that there are other phases with more complicated quasienergy spectra beyond $\theta_{c4}$. They can be studied straightforwardly using the analytical formula given above.

\begin{figure}
\includegraphics[width=0.4\textwidth]{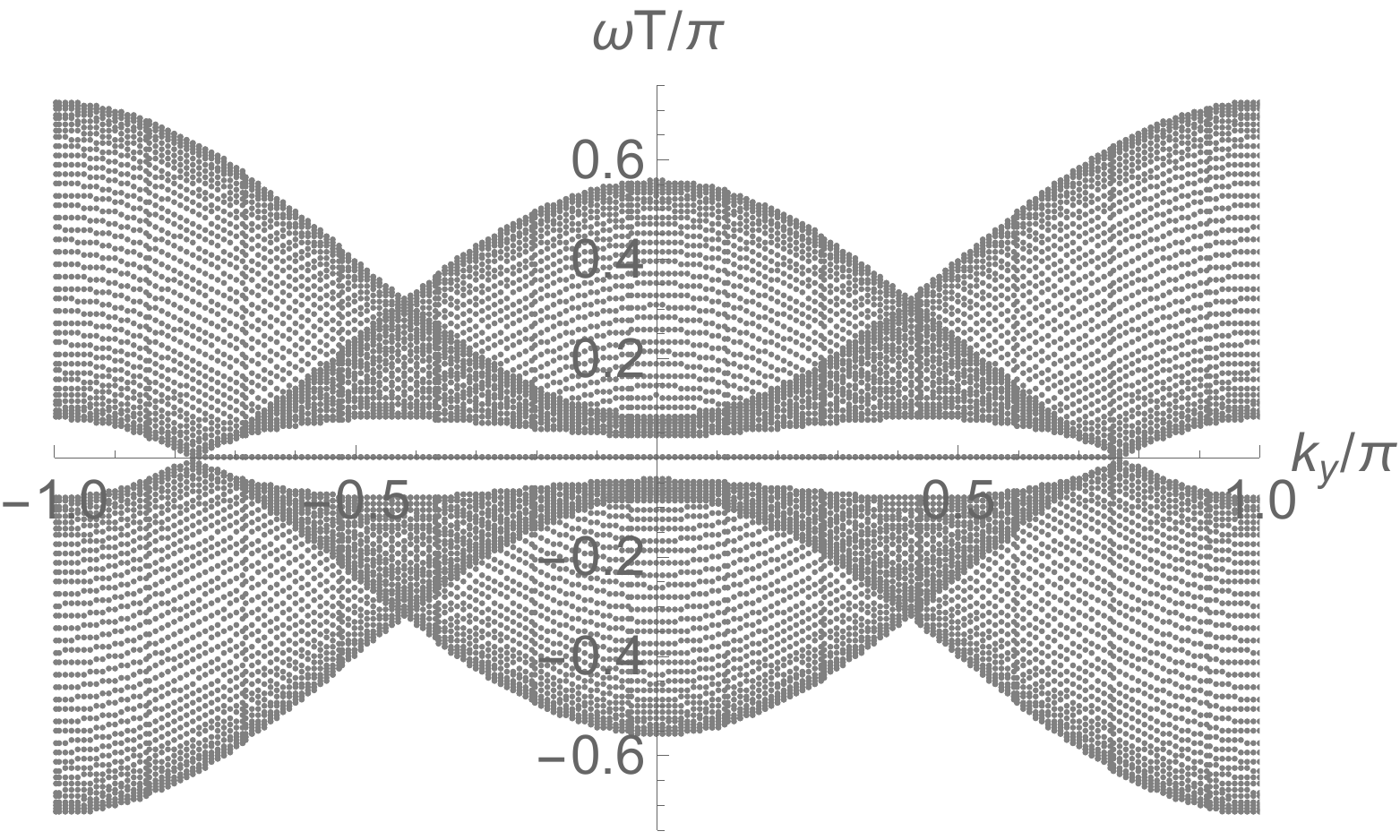}
\includegraphics[width=0.45\textwidth]{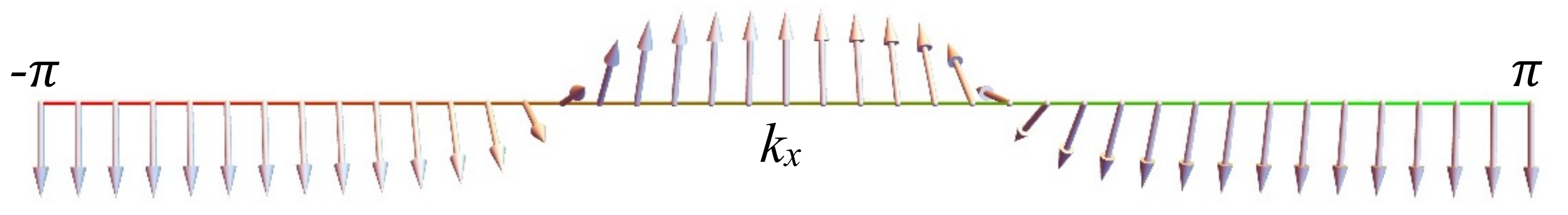}
\caption{The flat edge states for phase B, and the winding of the unit vector $\hat{d}(\mathbf{k})$ (the arrows) as $k_x$ is varied from $-\pi$ to $\pi$ for fixed $k_y=0.6\pi$. The winding number is $\mathsf{w}=-1$ for $k_y\in[-\kappa,\kappa]$. $\theta_y=0.5$, $\Delta=-\mu/2=0.25J$, $T J=0.5$.}
\label{fg2w1}
\end{figure}

\begin{figure}
\includegraphics[width=0.4\textwidth]{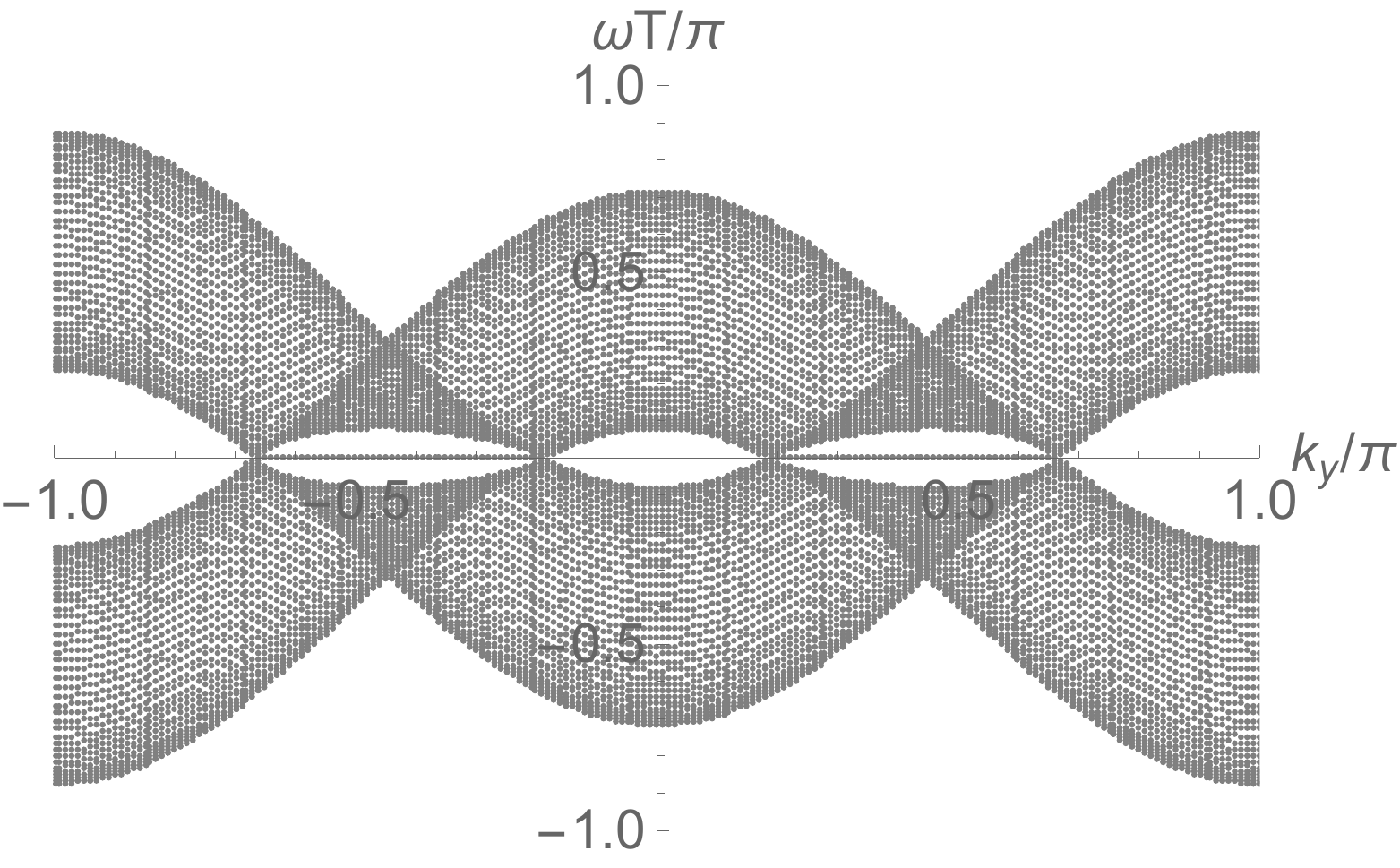}
\includegraphics[width=0.45\textwidth]{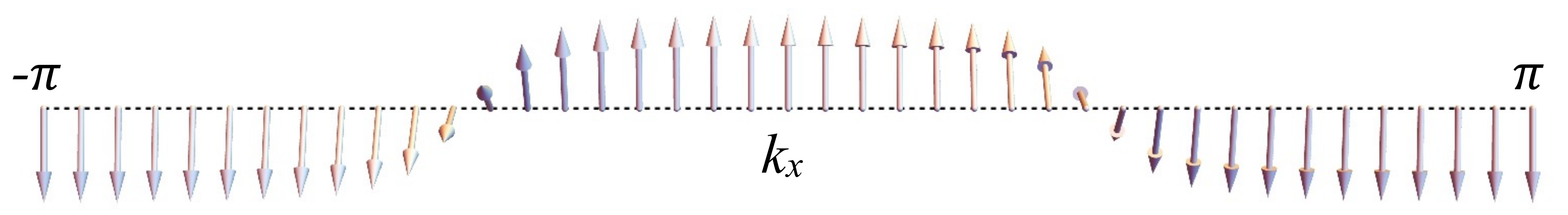}
\caption{The flat edge states of phase C, and the winding of $\hat{d}(\mathbf{k})$ for $k_y=0.5\pi$. The winding number $\mathsf{w}(k_y)=-1$ for $|k_y|\in [\kappa,\kappa']$. $\theta_y=0.75$, $\Delta=-\mu/2=0.25J$, $T J=0.5$.}
\label{fg2w2}
\end{figure}

\begin{figure}
\includegraphics[width=0.4\textwidth]{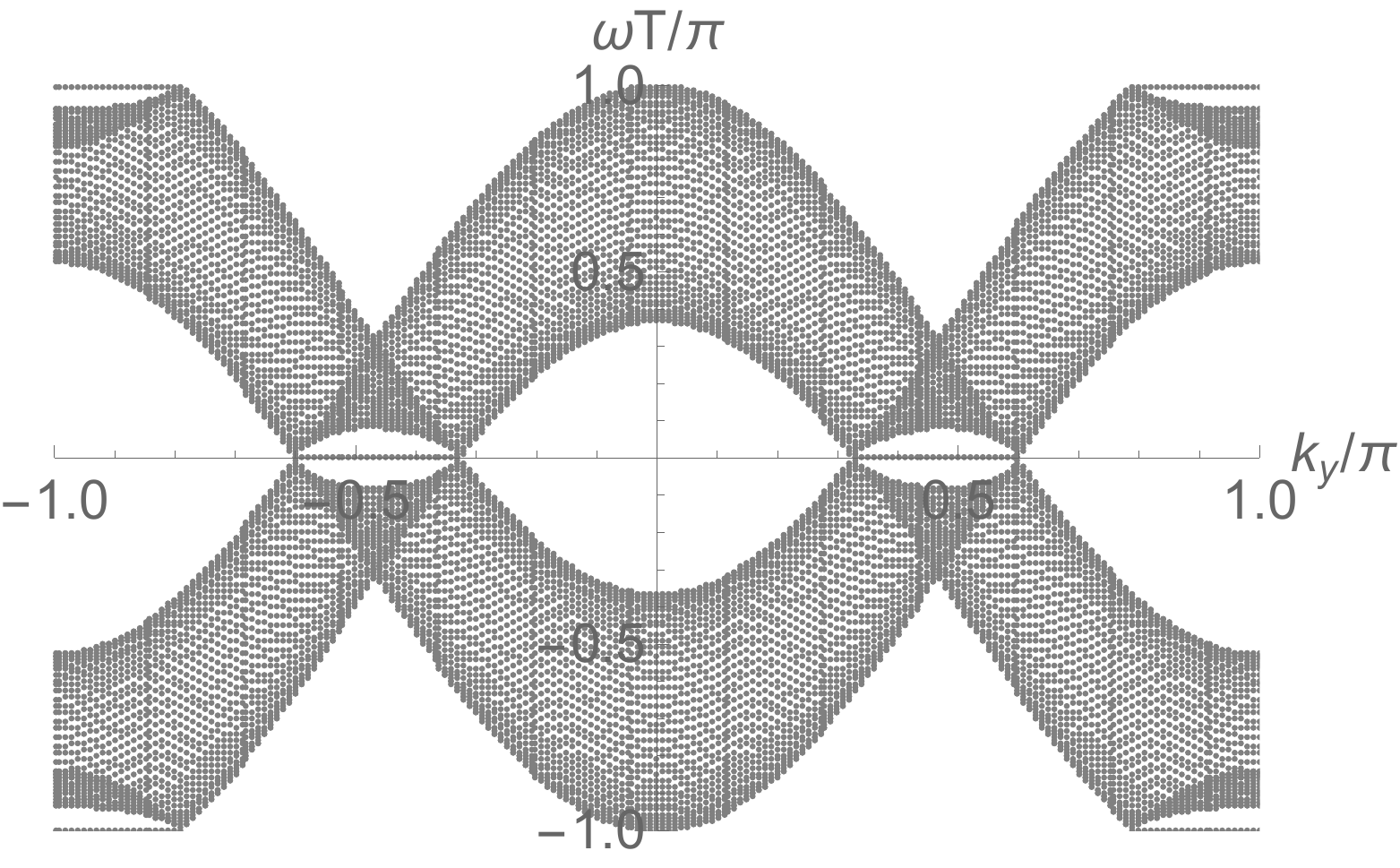}
\includegraphics[width=0.45\textwidth]{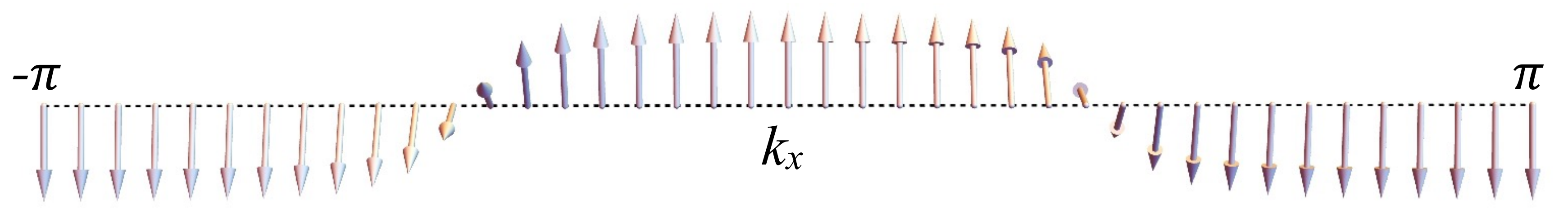}
\includegraphics[width=0.45\textwidth]{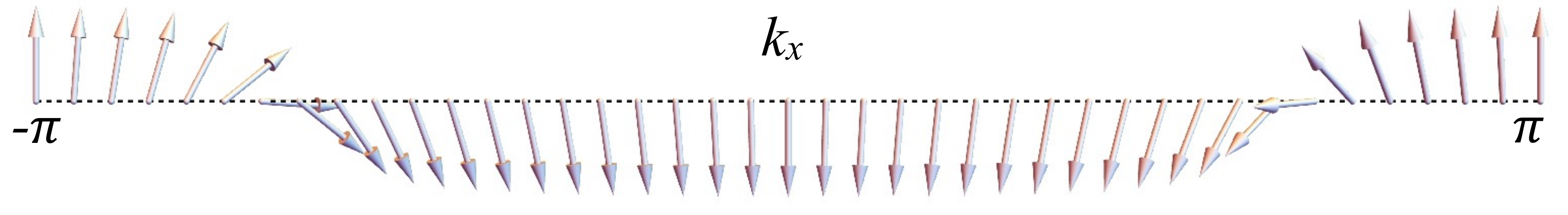}
\caption{The flat edge states for phase D at quasienergy zero and $\pi/T$ (top panel). The winding number for $\hat{d}(\mathbf{k})$ is $\mathsf{w}=-1$ for $k_y=0.5\pi$ (middle panel) and $k_y=0.9\pi$ (bottom panel). $\theta_y=1.2$, $\Delta=-\mu/2=0.25J$, $T J=0.5$.}
\label{fg2w3}
\end{figure}

All these gapless phases have nontrivial edge states. For completeness, we have included the representative spectrum of a finite slab with $L_x=40$ for phase B, C and D in Fig. \ref{fg2w1}, Fig. \ref{fg2w2}, Fig. \ref{fg2w3} respectively. Note that they are similar (but not identical) to the case of two-stage driving with finite $T-T_1$ presented earlier in Fig. \ref{fg2ed}. The edge states are all flat at either $\omega=0$ or $\omega=\pi$. Since $k_y$ is a good quantum number in the slab geometry, we can introduce a topological invariant known as the winding number $\mathsf{w}(k_y)$ for each $k_y$. It is defined as the number of times the vector $\mathbf{d}(\mathbf{k})$ winds (around the $x$ axis) as $k_x$ is varied from $-\pi$ to $\pi$ along a cut across the BZ. More precisely, let $\mathbf{d}_\perp=(d_y,d_z)=d_\perp(\cos\psi,\sin\psi)$,
\be
\mathsf{w}(k_y)=\frac{1}{2\pi} \int^\pi_{-\pi}\partial_{k_x} \psi (\mathbf{k})d k_x.
\ee
Note that $\mathsf{w}(k_y)$ defined here is unrelated to the winding number $w_\ell$ introduced earlier in section I. For phase B, $\mathsf{w}(k_y)=-1$ for $k_y\in [-\kappa, \kappa]$, as illustrated by the $\hat{d}$ vector in Fig. \ref{fg2w1}, and zero elsewhere. The value of $\mathsf{w}$ cannot change abruptly except at the nodal points where the gap closes and $\hat{d}$ becomes ill defined. Similarly, $\mathsf{w}(k_y)$ is nonzero for $|k_y|\in [\kappa, \kappa']$ in phase C and D. In all these cases, a finite (and constant) winding number $\mathsf{w}(k_y)$ guarantees a zero energy edge state for the corresponding $k_y$ value. The edge states stay flat at zero energy and terminate at the bulk nodal points. The bulk-boundary correspondence is identical to time-independent nodal superconductors as discussed in Refs. \cite{PhysRevB.83.224511,PhysRevB.84.060504} and Ref. \cite{PhysRevLett.89.077002} in a more general context.

The ``flat $\pi$-modes" in phase D is only possible in driven systems. Its existence can be understood as follows. Fig. \ref{fg2w3} shows that the winding number is $-1$ for $|k_y|\in [\kappa'',\pi]$. For these $k_y$ values, consider what has to happen at the edge where the sample ($\mathsf{w}=-1$) meets the vacuum (a trivial insulator with $\mathsf{w}=0$). At quasienergy $\pi/T$, on both sides of the edge the spectrum is gapped but endowed with different topological numbers. Thus the only way for $\mathsf{w}$ to change is the gap closing at $\pi/T$, i.e., the appearance of a $\pi$-mode at the edge. The constancy of $\mathsf{w}$ within the interval $[\kappa'',\pi]$ also dictates that the $\pi$-modes have to stay flat and terminate at the quasienergy degeneracy points at $k_y=\pm\kappa''$. Therefore, the appearance of the flat $\pi$-modes can be traced back to the degeneracies of the quasienergy bands at the QBZ boundary. 

In summary, under the two-step driving, the dynamically coupled Kitaev chains can become a gapless Floquet superconductor with isolated point nodes and flat edge states. This includes phase B which has one pair of points nodes and a line of edge states at zero energy (Fig. \ref{fg2w1}), phase C with two pairs of point nodes (Fig. \ref{fg2w2}), and phase D with flat edge states at quasienergy zero as well as $\pi/T$ (Fig. \ref{fg2w3}). While the model itself is easy to solve numerically, our analysis has focused on simplifying the algebra in the limit of periodic kicking to clearly define the series of phases via their nodal structures, edge states, and topological invariants.

\section{Final remarks}

We end our discussion with a brief comment on the possibility of realizing the model of dynamically coupled Kitaev chains proposed and analyzed here. The choice of the model is largely motivated by its simplicity (the physics of a single chain is thoroughly understood) as well as its potential connection to other known topological $p$-wave superconductors. Dipolar Fermi gases of magnetic atoms (Er \cite{Aikawa19092014} and Dy \cite{PhysRevLett.108.215301}) or dipolar molecules (KRb \cite{ni} or NaK \cite{PhysRevLett.114.205302}) have been shown theoretically to support $p$-wave pairing due to attractive part of the dipole-dipole interaction \cite{Baranov-rev}. For dipolar fermions loaded on a square optical lattice,
tilting the dipoles along the $x$-direction by an external field will result in an attractive interaction between fermions on two neighboring sites along $x$ but a repulsive interaction along $y$. On the mean field level, it supports $p_x$ pairing \cite{PhysRevLett.108.145301} and thus the term $\Delta c^\dagger_{\mathbf{r}} c_{\mathbf{r}+\hat{x}}$. If we further treat the repulsive interaction on the Hartree-Fock level, its dominant effect  is to renormalize the chemical potential. The optical lattice can be modulated dynamically to allow or shut off the hopping along $x$ or $y$.
And the result is a system approximately described by the dynamically coupled Kitaev chain model. In practice, the implementation of the model may not be easy. As a toy model, its main virtue is to show us what dynamical driving is capable of: the transmutation of the edge modes, from Majoranas at zero energy to linearly dispersing chiral fermions or flat $\pi$-modes, is perhaps as dramatic as it could be. 

\acknowledgments
This work is supported by AFOSR grant number FA9550-16-1-0006 and NSF grant number PHY-1205504. The author is grateful to Ludwig Mathey and the Center for Optical Quantum Technologies at University of Hamburg for hospitality and support during the workshop ``Emergence in driven solid-state and cold-atom systems." Stimulating discussions with Ahmet Keles, Lijun Lang, Takashi Oka, Indu Satija, and Qi Zhang on related topics are greatly acknowledged.

\bibliography{coldatoms,driving}

\end{document}